\documentclass[aps,prb,twocolumn,floatfix,superscriptaddress,10pt]{revtex4}

\usepackage{graphicx}
\usepackage{amsmath}
\usepackage{amsfonts}
\usepackage{amssymb}

\usepackage[T1]{fontenc}
\usepackage[cp1250]{inputenc}

\begin{document}

\bibliographystyle{unsrt}

%\linespread{1.3}
%\setlength{\oddsidemargin}{-1.0cm} \setlength{\textwidth}{17.2cm}
%\setlength{\textwidth}{470pt}

\newcommand{\figwidth}{0.9\columnwidth}
\newcommand{\hc}{\text{H.c.}}

\title{Spin rotational symmetry breaking by orbital current patterns in two-leg ladders}
\author{P. Chudzinski}
\affiliation{DPMC-MaNEP, University of Geneva, 24 Quai
Ernest-Ansermet CH-1211 Geneva, Switzerland}
\author{M. Gabay}
\address{Laboratoire de Physique des Solides, Bat. 510,
Universit\'e Paris-Sud 11, Centre d'Orsay, 91405 Orsay Cedex,
France}
\author{T. Giamarchi}
\affiliation{DPMC-MaNEP, University of Geneva, 24 Quai
Ernest-Ansermet CH-1211 Geneva, Switzerland}

\begin{abstract}
 We investigate the physical consequences of orbital
current patterns (OCP) in doped two-leg Cu-O Hubbard ladders.
The internal symmetry of the pattern,  in the case of the ladder structure,
differs slightly from that suggested so far for
cuprates. We focus on this OCP and look for measurable signatures
of its existence. We compute the magnetic field produced by the  OCP at each
lattice site, and estimate its value in view of a possible experimental detection. Using a renormalization group (RG)
analysis, we determine the changes that are caused by the SU(2)
spin-rotational symmetry breaking which occurs when the OCP is present in the ground
state phase diagram. The most signifcant one is an
in-plane SDW gap opening in an otherwise critical phase, at intermediate
dopings. We estimate the value of this gap, give an
analytic expression for the correlation functions and examine some of the
magnetic properties of this new phase which can be revealed in measurements. We compute
the conductance in the presence of a single impurity, using an
RG analysis. A discussion of the various sources of SU(2)
symmetry breaking underscores the specificity of the OCP induced
effects.
\end{abstract}

\maketitle

\section{Introduction}

More than two decades after the experimental discovery of a
pseudogap in the phase diagram of high temperature
superconductors (HTSC) \cite{Alloul_YBCO}, a microscopic
explanation of its origin still remains elusive. Many scenarios
have been proposed \cite{PALeeetal_review,
lederer_superconductivity_flux_ref,
kotliar_liu_dwave_slavebosons, chakravarty_ddw_pseudogap,
affleck_marston, senthil_09, gabay_RMP, Kivelson_RMP} but so
far no consensus has emerged on any particular one. Among the
contenders are proposals that the pseudogap might be linked
to current formation. It was suggested that staggered current patterns (DDW phase)
\cite{chakravarty_ddw_pseudogap} would lead to
a pseudogap due to the doubling of the unit cell. An alternative uniform current pattern\cite{varma_3band_model,
CVarma_orbitalcurrents} was proposed by Varma. In this latter case the three band
nature of the system is crucial, since currents flow along
closed loops formed by $Cu$-$O$ bonds in such a way that
translational symmetry is preserved, while time-reversal
symmetry is broken. On the experimental front specific
signatures compatible with uniform currents where
observed\cite{BFauque_neutrons, Xia_optOAF, Kapitulnik_OCP,
Baledent_OCP} even if the issue is still controversial
\cite{Sonier_OCPneutron, strassle_lackOCP}. On the theoretical
side significant efforts were made to probe the possible
existence of such a phase, either using numerical approaches
\cite{greiter_exactdiag_currents,thomale_exactdiag_currents,weber_OCP} 
or starting from Mott insulating side \cite{Bulaevskii_added}.

As a way of investigating the physics of these states in a more controlled
way and also in view of possible connections with experimental
systems\cite{piskunov_ladder_nmr,
piskunov04_sr14cu24o41_nmr,fujiwara03_ladder_supra,
imai_NMR_doped_2ladder,kumagai_NMR_2ladder},  phases exhibiting
OCP were also considered in the context of two-leg
ladders\cite{orignac_2chain_long, schollwock_CDW+current}.
Indeed for this essentially one-dimensional (1D) case, various
analytical tools allow one to explore the consequences of such an
unusual phase. In previous work\cite{chudzinski_ladder_long,
chudzinski_ladder_rapid} we established that in the weak
coupling limit of $Cu$-$O$ Hubbard ladders there exists a range
of doping with a quasi-long range ordering of orbital currents, such
that the pattern is symmetric with respect to the exchange of
the two legs (\emph{o}-OCP). In these studies, however,
possible changes in the symmetry of the system caused by the 
formation of the orbital currents  was not taken into account. Our analysis of the two-leg
ladder problem was based on the assumption that the spin
degrees of freedom were totally decoupled from the motions of
the carriers in real space. Although this is a very good
starting point when the order parameter is small, it is
interesting to investigate the consequences of such a coupling.
In particular in the presence of a static pattern of currents,
it is natural to expect that the spin of the electrons will
couple to the generated moment, and lead to interesting
spin-orbit effects.

The aim of the present paper is to explore and to discuss 
physical effects which are occurring in response to the
\emph{o}-OCP. Anticipating the results presented below, the
main consequence of the \emph{o}-OCP is the appearance of
magnetic moments which break SU(2) spin-rotational symmetry.
While measuring the primary (first order) response of the
system to the \emph{o}-OCP is expected to be experimentally
challenging, these collateral effects constitute clear
fingerprints of the existence of such a phase. We also check
the stability of the \emph{o}-OCP itself with respect to the
second order perturbation which couples the order parameter
with the spin of the electrons. We argue that SU(2) symmetry is
fairly well protected (in the absence of crystalline
anisotropies) and that its breaking is a telltale of the
existence of the \emph{o}-OCP.

The paper is organized as follows: At the beginning of Sec.
\ref{sec:model} we introduce the Hubbard model for two-leg
$Cu$-$O$ ladders and we review the main results that were
obtained for this model in the weak coupling renormalization
group (RG) approach. Next we derive the additional terms that
arise in the presence of the \emph{o}-OCP. The magnetic moments
that they induce in the ladder are calculated and the first
order effect that they generate are also discussed there. New
terms in the hamiltonian will be derived. Sec.
\ref{sec:results} is devoted to an RG analysis of the
generalized ladder model obtained in the previous section.
Experimental consequences entailed by these results are
described in Sec. \ref{sec:discussion}. Lastly, in the
Appendix, we discuss other possible sources of SU(2) symmetry
breaking in two-leg $Cu$-$O$ ladders (such as Rashba effect).

\section{The model}\label{sec:model}

Let us begin with a brief review of two-leg $Cu$-$O$ ladder
physics. We consider a two-leg $Cu$-$O$ ladder with local
interactions. The Hamiltonian of this system contains two parts:
the kinetic energy of electrons moving on the lattice $H_{T}$ and
electron interactions $H_{int}$
\begin{equation}\label{hubb}
    H=H_{T}+H_{int}
\end{equation}
In the SU(2) invariant case, the physics of this hamiltonian is
known (assuming that $H_{int}$ is treated as a perturbation of the
kinetic term)
\cite{chudzinski_ladder_rapid,chudzinski_ladder_long}. In the
first part of this section, we recall its main characteristics, we
show the condition required for the \emph{o}-OCP to exist, and we
also introduce relevant notations. In the second part, we explain
how the \emph{o}-OCP breaks spin-rotational symmetry and we derive
new terms in the hamiltonian that arise because of the broken
symmetry. In principle, these terms affect both $H_{T}$ and
$H_{int}$. This secondary effect is caused by the presence of the
\emph{o}-OCP.

\subsection{The $Cu$-$O$ Hubbard ladder}\label{cleanlad}

Including two $Cu$ and five $O$ atoms in the unit cell, the
tight-binding kinetic energy part $H_{T}$ reads
\begin{widetext}
\begin{multline}\label{hubbT}
    H_{T}=\sum_{j\sigma}( \sum_{m\in Cu}\epsilon_{Cu}
    n_{mj\sigma}+ \sum_{m\in O}\epsilon_{O}
    n_{mj\sigma}- \sum_{m\in Cu} t
    [a_{mj\sigma}^{\dag}(b_{mj\sigma}+b_{mj-1,\sigma})+\hc] \\
- \sum_{m\in Cu} t_{\bot}
    [a_{mj\sigma}^{\dag}(b_{m+1,j\sigma}+b_{m-1,j\sigma})+\hc])
-\sum_{m\in O(leg)} t_{pp}
    [b_{mj\sigma}^{\dag}(b_{m+1,j\sigma}+b_{m-1,j\sigma}+b_{m+1,j-1\sigma}+b_{m-1,j-1\sigma})+\hc])
\end{multline}
\end{widetext}
where $a_{mj\sigma}(b_{mj\sigma})$ is the creation operator of
holes with spin $\sigma$ on a copper (oxygen) site (\emph{j} is a
site along chain and \emph{m} labels the atoms in each cell);
$n_{mj\sigma}^{Cu}=a_{mj\sigma}^{\dag}\cdot a_{mj\sigma}$ We use
hole notation such that $t$, $t_{\bot}$,$t_{pp}$ are all positive.
$\epsilon=\epsilon_{O}-\epsilon_{Cu}$ is the difference between
the oxygen and copper on-site energies.

The model is reduced to two low lying bands crossing the
Fermi energy. They are denoted $o$ (symmetric under the exchange
of the two legs) and $\pi$ (antisymmetric under the exchange of
the two legs). With $\alpha$ ($=0,\pi$) and $\sigma$ denoting the
band and the spin index respectively, the Hamiltonian reads
\begin{equation}\label{HTdiag}
    H_{T}=\sum_{k\alpha\sigma}e_{\alpha}(k)n_{k\alpha\sigma}
\end{equation}
where
\begin{equation}\label{bazy1}
    a_{mk\sigma}= \sum_{\alpha} \lambda_{m\alpha} a_{\alpha k\sigma}
\end{equation}
and $e_{\alpha}(k)$ and $\lambda_{m\alpha}$ are the eigenvalues
and components of the eigenvectors of the Hamiltonian matrix (see
Ref.~\onlinecite{chudzinski_ladder_long}).

In the low energy limit one may linearize the dispersion relation
in the vicinity of the Fermi energy:
\begin{equation}\label{contin}
    H_{T}=\sum_{|q|<Q} \sum_{r\alpha\sigma}r q V_{F\alpha} a^{\dag}_{\alpha r
    q\sigma} a_{\alpha r q\sigma}
\end{equation}
and it is easy to bosonize this free fermion theory
\cite{giamarchi_book_1d,voit_bosonization_revue}. Two charge (c)
and spin (s) boson phase fields $\phi(x)$
are introduced for each fermion species
($x$ is the spatial coordinate along the ladder).

We also introduce the phase fields $\theta$; their spatial
derivative $\Pi(x)=\pi\nabla\theta(x)$ is canonically conjugated
to $\phi(x)$. Now the Hamiltonian may be rewritten using the above
phase fields. The kinetic part and those pieces in $H_{int}$ which
can be expressed as density-density terms give rise to the
following quadratic form
\begin{equation}\label{eq:Hbozon}
    H_{0}= \sum_{\lambda} \int \frac{dx}{2\pi}[(u_{\lambda}K_{\lambda})(\pi \Pi_{\lambda})^{2}+(\frac{u_{\lambda}}{K_{\lambda}})(\partial_{x} \phi_{\lambda})^{2}]
\end{equation}
in the diagonal basis $B_0$. $\lambda=1,2,3,4$ labels the
eigenmodes ($1,2$ are spin modes and $3,4$ are charge modes
\cite{chudzinski_ladder_long}) For the non-interacting system
one has $K_{\lambda}=1$ for all modes; the diagonal density
basis is then the bonding/antibonding one $B_{o\pi}$ (the
momentum $k_{\perp}$ associated with the rungs is either 0 or
$\pi$). The other basis which is commonly used in the
literature is the total/transverse one, $B_{+-}$. It is related
to $B_{o\pi}$ by:
\begin{equation}\label{eq:basis}
    \phi_{\nu+(-)}=\frac{\phi_{\nu o}\pm\phi_{\nu \pi}}{\sqrt{2}}
\end{equation}
where $\nu =$c or s, depending on
which particular density one considers.

\begin{figure}[h]
  \centering
  \includegraphics[width=\figwidth]{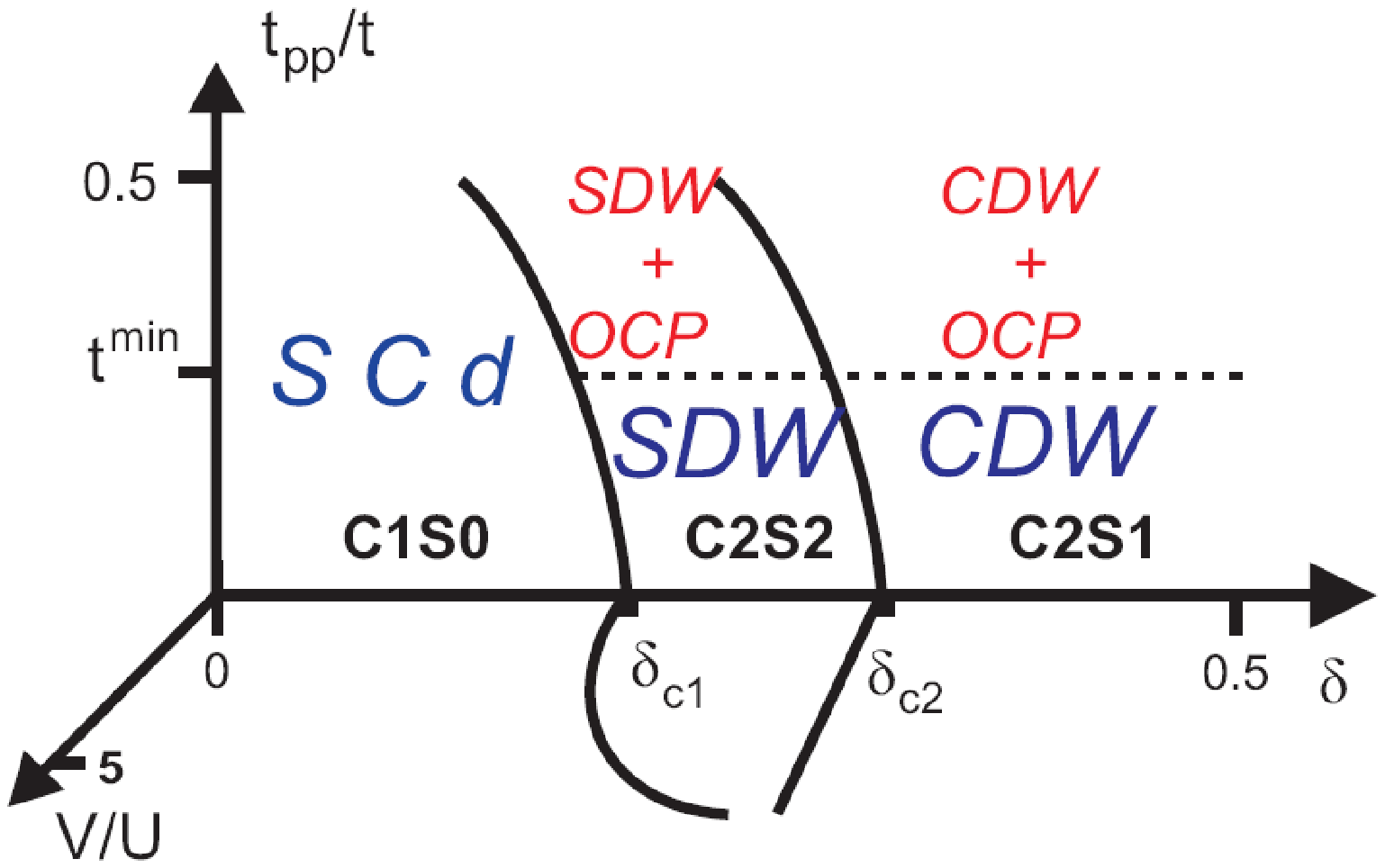}
  \caption{The phase diagram of two-leg $Cu$-$O$ Hubbard ladders versus doping
  for $U_{Cu}>0$. Zero doping corresponds to
  the half filled case. Umklapp terms which open a gap in
  the charge symmetric mode are not included here \cite{chudzinski_ladder_long}.
   SCd, CDW and SDW denote respectively d-wave superconductivity, charge-density
  wave and spin-density wave states. OCP indicates that an orbital current
  pattern is present on top of a density wave instability}\label{fig:dopin}.
\end{figure}

The interaction part, in fermionic language, is given by
\begin{eqnarray}
  H_{int} =\sum_{j} (\sum_{m\in Cu} U_{Cu}n_{mj\uparrow}n_{mj\downarrow} + \sum_{m\in O}U_{O} n_{mj\uparrow}n_{mj\downarrow}\nonumber \\
 + \sum_{m\in Cu, n\in O}\sum_{\sigma, \sigma'}V_{Cu-O}n_{mj\sigma}n_{nj\sigma'}) \label{hubbInt}
\end{eqnarray}
Eq.~(\ref{hubbInt}) gives rise to two types of terms: the first
ones are of the forward scattering type and they can be cast in a
quadratic form
$(\sim \nabla\phi(x))^{2}$ in bosonization language. %$(\nabla\phi(x))^{2}$.
These terms are then incorporated in the above mentioned matrix
$\hat{K}$ of Luttinger liquid (LL) parameters, and hence they are
treated exactly in this procedure. Since the form of $\hat{K}$
depends on the basis in which the densities are expressed, the
Hamiltonian will take the simple form Eq.~(\ref{eq:Hbozon}) in the
eigenbasis of the matrix $\hat{K}$.
% if we express $\hat{K}$ in its eigenbasis.
The remaining interaction terms yield non-linear cosines and these
are the ones for which the RG procedure is required. In Refs.
~\onlinecite{chudzinski_ladder_rapid,chudzinski_ladder_long} we showed
that the eigenbases for the spin and for the charge modes rotate
during the RG flow. Two fixed points were found, namely $B_{+-}$
and $B_{o\pi}$. Interband physics dominates in the former case
(mixing of the $o$ and $\pi$ bands) and intraband physics, in the
latter case. Close to half filling (low doping regime), $B_{+-}$
is the fixed point basis for the spin and the charge modes. Both
spin modes are gapped, and so is one of the charge modes. Using
the notation of Balents and Fisher \cite{balents_2ch} the ladder
is in the \emph{C1S0} state (\emph{CnSm} denotes $n$ ($m$) gapless
charge (\emph{C}) (spin \emph{S}) modes). For an intermediate
range of dopings, $B_{+-}$ is the fixed point basis for the spin
variables and $B_{o\pi}$ the fixed point basis for the charge
variables. This is the \emph{C2S2} regime when all modes are
gapless. The decoupling of the spin and of the charge eigenbasis
is responsible for this quantum critical state. For higher
dopings, $B_{o\pi}$ is the fixed point basis for the spin and the
charge modes. The ladder is in the \emph{C2S1} phase where the $o$
spin mode is gapped. The corresponding phase diagram, which was
established in
Refs.~\onlinecite{chudzinski_ladder_rapid,chudzinski_ladder_long}, is
shown in Fig.\ref{fig:dopin}.

A noteworthy result that came out of our analysis was that the
symmetry of the OCP is that of the o band, for the $Cu$-$O$
ladder. What is more, there exists a range of values of $t_{pp}$,
above some critical hopping where this phase is dominant. In the
following, we present a detailed discussion of experimentally
accessible effects caused by the presence of this unusual phase.

\subsection{Primary effect induced by the \emph{o}-OCP: magnetic moments}

When the \emph{o}-OCP is present, its current pattern $I(x)$ generates a
local magnetic field $B({\mathbf x})$ at each site {\bf x} of the
lattice. This is the quantity that we are interested in. The value
of the field can be obtained from the Biot-Savart law
\begin{equation}\label{eq:fieldB}
    B({\mathbf x})=\frac{\mu_{0}}{4\pi}\oint I(x)\frac{d{\mathbf l}\times {\mathbf r}}{r^3}=\frac{\mu_{0}}{4\pi}\sum_{n}I_{n}\oint'_{n}\frac{d{\mathbf l}\times {\mathbf r}}{r^3}
\end{equation}
${\mathbf r}$ is the vector joining the center of an infinitesimal
element of current $ I d{\mathbf l}$ to the lattice site ${\mathbf
x}$. In the second equality, the prime on the integral means that
we are integrating over an elementary $Cu$-$O$ loop (triangle) --
denoted by $n$ -- on the ladder; we sum over all triangles and we
use the fact that current is conserved for each loop $n$ of the
lattice.

To perform the summation in Eq.~(\ref{eq:fieldB}), we first single
out those triangles that belong to the particular unit cell that
contains the lattice site {\bf x};
we immediately notice that we get a zero net contribution whenever
\begin{itemize}
    \item the current flows in a direction that passes through the lattice site {\bf x}
    (the cross product $d{\mathbf l}\times {\mathbf r}$ gives zero)
    \item the high symmetry of the OCP causes a cancellation of the fields due to
    currents connected by mirror symmetry (for that reason, the field on the on-rung oxygen is equal to zero)
\end{itemize}
These properties can easily be accounted for if one rewrites
the magnetic field using a multipole expansion of the vector potential
\begin{equation}\label{eq:Bmulitip}
    A(x)= \frac{\mu_{0}}{4\pi}\sum_{n}I_{n}\sum_{m}\frac{1}{r_{n}^{m+1}}\oint (r')^{m}P_{m}(\cos\bar{\theta})dl
\end{equation}
There are no magnetic monopoles, dipole contributions vanish
because of the cancellations described above and so do quadrupole
contributions inside each cell. Symmetry allows us to conclude
that octupoles are the lowest order non-vanishing terms. The above
formula can be significantly simplified thanks to the fact that
all atoms lie in the same plane from which we have
$\cos\bar{\theta}=1$; we can also assume that the approximate
distance between the moments $r'$ is equal to unit cell size $a$.

Thus, we notice that we may get an accurate estimate of the
contribution of all the other cells, if we assimilate them to
octupole moments. Basically this approximation amounts to
replacing the primed integrals within each triangle by a number
(the magnetic moment at the center of each elementary triangle).
The current creates a dipole moment perpendicular to the plane of
the triangle (and proportional to the current flowing around it).
Four neighboring current triangles, centered at a single,
\emph{n-th} $Cu$ site, define an elementary octupole $\zeta(n)$
(with a shape of a square). Then the magnetic field is
approximated as a sum of octupoles along the ladder:
\begin{equation}\label{sum:octup}
    B= \frac{\mu_{0}}{4\pi}[\frac{\zeta(x)}{a^{5}}+\sum_{n}
    (\frac{\zeta(n)-\zeta(-n)}{r(x_{n})^{5}}+\frac{\zeta(n)-\zeta(-n)}{(r(x_{n})^{2}+a^{2})^{5/2}})]
\end{equation}
The first term comes from octupoles on opposite legs, the sum runs
along the rungs of the ladder, with the first piece arising from
same leg contributions and the second piece from opposite leg
contributions. A crucial point is how to estimate the value of
$\zeta(n)\sim I_{n}$. This work pertains to physical systems quite
similar to cuprates. Experimental magnitudes of the local magnetic
(dipolar) moments, which are tilted $45^{\circ}$ out of the
$Cu$-$O$ plane, are of order $0.1\mu_{B}$ \cite{BFauque_neutrons}.
Thus we assume that our maximal octupolar momentum (on top of the
density wave (DW)) is equal to $\zeta(I_{max})\simeq
4\frac{\sqrt{2}}{2}0.1\mu_{B}a^{2}$ and we will use that value in
our estimates.
The case of on-leg oxygens, which lie at the boundary of each
cell, is special because then the quadrupolar contribution does
not vanish, and it will be discussed below.

One has to remember that the periodicity along the ladder plays an
important role in the computation of such a sum; specifically, if
the \emph{o}-OCP were uniform along the ladder, all the different
contributions would cancel out, because then $\zeta(n)=\zeta(-n)$,
and the total magnetic field would be zero on each atom. In our
case we have an \emph{o}-OCP on top of a DW with a real space
periodicity $(2k_{Fo})^{-1}$, hence a magnetic field with the same
periodicity appears.

\begin{figure}[ht]
  \centering
  \includegraphics[width=0.45\textwidth]{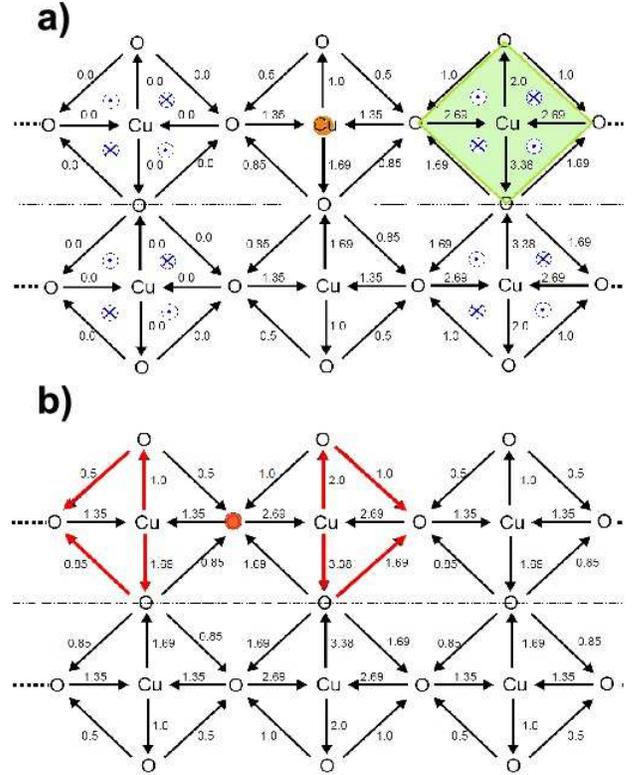}\\
  \caption{The two main contributions to the amplitude of the magnetic field at
   lattice site {\bf x} (orange circle) as a result of the \emph{o}-OCP flow:
  (a) the octupole moments (single octupole = green square) which should be summed up for all elementary
  unit (squares) along the ladder (except the one which contains lattice site {\bf x}) and
  (b) the quadrupole magnetic moment that is present only if {\bf x}
  is one of the on-leg oxygens; the field comes from the asymmetry of OCP amplitudes in two
  neighboring unit cells (non-zero contribution indicated in red); crosses (dots) correspond to single
  magnetic dipoles pointing into (out of) the page}\label{fig:fields}
\end{figure}

As a result one finds that, depending on the nature of the lattice
site {\bf x}, the magnetic field contains either only octupolar or
both quadrupolar and octupolar contributions.
\begin{itemize}
    \item for each on-leg oxygen atom, there are quadrupolar contributions
    coming from the currents flowing along the neighboring elementary
    cells which are not pointing out to this atom (the intensities of the
    currents are different in the two neighboring cells); see Fig.\ref{fig:fields} b)
    numerically,  we get a  value of $50[Oe]$ for the field.
    \item for all other atoms (except the on-rung oxygen) we get
    octupolar contributions (as shown in Fig.\ref{fig:fields} a)) coming from currents flowing through unit cells  other than that of the atom:
    one piece stems from the single octupole on the leg opposite to where site {\bf x} resides  and the other
%    the opposite leg than site {\bf x} and
from a sum of double octupoles along the ladder (a series oscillating with a $(2k_{Fo})^{-1}$
periodicity and decreasing as $1/r^{5}$); numerically,  we get a  value of $10\div 20[Oe]$ for the field. % this contribution is smaller than the first
\end{itemize}
We note that both types of contributions produce a staggered field
pointing in a direction perpendicular to the plane of the ladder,
which possesses the periodicity of the \emph{o}-OCP. The dipolar
component and one of the terms in the octupolar series are
illustrated in the figure below.

This local magnetic field $\vec{B}(x)$ break SU(2) spin-rotational
symmetry. Usually, such symmetry breaking may occur in the presence of
spin-orbit, which adds a $\vec{L}_{eff}(x)\cdot \vec{S}(x)$
term to our Hamiltonian (it should be emphasized that both
vectors are taken at the same point). Tools required to deal
with such a perturbation are readily available. We simply need
to tailor these to our case. Both operators are two
\emph{o}-band fermion combinations, which, when coupled, will
give four fermion products (to first order). Detailed
analysis shows that such $\vec{L}_{eff}(x)\cdot \vec{S}(x)$ terms
indeed cover the presence of magnetic field
provided that hypothetical (periodic along x-axis) field $\vec{B}(x)$
would have had only z-component. For example we may check the form (in bosonization language) of
four fermion operators describing the coupling of the SDW with the spin degree
of freedom of carriers (an example of this type of coupling is:
$\psi_{oL\uparrow}^{\dag}(x)\psi_{oR\uparrow}(x)(\psi_{oL/R,\sigma}^{\dag}(x)\hat{\sigma}\psi_{oL/R,\sigma}(x))$)
Thus, one can use the theory of spin orbit coupling (as a perturbation) to solve our
problem. The z-axis orientation of our perturbation is a
consequence of the high symmetry of the lattice, a point which we
emphasize in the next section.

\subsection{Side effect of the \emph{o}-OCP: new terms in the hamiltonian}

\subsubsection{First order term: spin dependent hopping}\label{subsec:spindependenthopping}

Because of the local magnetic fields that are generated by the
\emph{o}-OCP, the on-site energies of each atom (except for the
on-rung oxygens), will be different for carriers with spin up
and spin down. This can be viewed as local Zeeman splittings,
which also affect overlap integrals for different spin
rotations. The d-p orbital overlaps, and thus the hopping
amplitudes, will depend on $S_{z}$. This suggests that
spin-dependent hopping will be present. The additional hopping
term is defined by $t_{\sigma\sigma'}=\imath
\vec{t}\cdot\vec{\sigma}$, where $\vec{\sigma}$ are Pauli
matrices. Our argument implies that $\vec{t}$ has only
z-components. In general there can be also other sources of
$\vec{t}$ like coupling of electrons orbital momentum $\hat{L}$
with magnetic moment of each triangle.

For $Cu$-$O$ lattices, we can make generic statements regarding
the possible spin-dependent hoppings caused by interactions with
finite out of plane moments (the treatment is similar to a crystal
field splitting with a specific direction in space):
\begin{equation}\label{eq:hopp}
    \vec{t}=\imath \lambda(B)\sum_{n}\frac{\langle d_{x^{2}-y^{2}}|\hat{\vec{L}}|d_{n} \rangle}{\epsilon_{x^{2}-y^{2}}-\epsilon_{n}} t_{p_{\gamma}d_{n}}
\end{equation}
Here $d_{x^{2}-y^{2}}$ is the ground state d-orbital of the $Cu$
atom located at some site $i$; $|d_{n}\rangle$ is an excited $d$
state of that atom; $t_{p_{\gamma}d_{n}}$ is the hopping amplitude
between the $Cu$ atom in the state $|d_{n}\rangle$ and a
neighboring $O$ atom in one of the $p$ orbitals represented by the
state $|p_{\gamma}\rangle$; $\hat{\vec{L}}$ is the orbital moment.
One has to remember that symmetry requires some of the hopping
integrals to vanish (for instance $t_{p_{\gamma}d_{xz}}=0$ and
similarly $\hat{L}|d_{3z^2-r^2}\rangle=0$). Thus lattice symmetry
implies that $\vec{t}$ has always only a z-component.

Taking this extra term into consideration, first order effects can
be easily incorporated in our tight binding model: the problem can
still be factorized in the spin variables ($\vec{t}$ has only a z
component). The magnetic field has \emph{o}-band wave-function
symmetry, thus Zeeman splittings (and also overlap integrals
variations) will have the same intra-cell symmetry. For the OCP
case, the splitting is purely vertical and it oscillates with
$k_{Fo}$ periodicity from unit cell to unit cell (whereas in the
more frequently discussed Rashba case it is horizontal and the
same for all the unit cells), which gives constructive
interference within that band. The result is that the
\emph{o}-band is split -- the hopping parameters of spin up and
down are slightly different.

There are three consequences of this type of splitting:
\begin{itemize}
    \item the Fermi velocities of spin up and down states are not equal anymore
    \item there is a mismatch of their Fermi vectors
    \item the two bands may contain different amounts of $d$ and $p$ orbitals
\end{itemize}
We do not need to discuss the last item, because the
eigenvectors of bands pairs are very similar. Let us instead
focus on the first two. The difference in the velocities,
translated into boson field language, yields terms of the form
$\phi_{c +}\Pi_{s +}$, which affect the values of the
parameters $K_{c,s+}$ and the form of correlation functions.
The modification of $K_{c,s+}$ has to be taken into account in
the initial condition of the RG flow;  the impact on
correlation functions will be described in detail in
Sec.~\ref{sec:discussion}. The Fermi vector mismatch has an
adverse impact on perpendicular (OCP case) spin scattering,
because momentum is not conserved during that process, any
longer. This effect will be accounted for much in the same way
as one treats slight departures from commensurability with
doping, in Mott transitions.

In our case, we may estimate $|\vec{t}|/t < 10^{-4}$ (considering that the
 magnetic moment of a single loop is $\lambda\sim
0.1\mu_{B}$). This value gives $\delta k_{Fi}/k_{Fi}$ of the same
order, so all these effects can indeed be treated as a
perturbations.

\subsubsection{Second order terms: spin-flip scattering channels}

Let us turn to the next order in perturbation, which yields four
fermion operators. In general these can be written in a
Dzyaloshinskii-Moryia (D-M) form
\begin{equation}\label{eq:D-M ham}
    H_{D-M}=\vec{D_{ij}}\cdot(\vec{S_{i}}\times \vec{S_{j}})
\end{equation}

The value of the D-M interaction between two $Cu$ sites $i$ and
$j$ separated by an $O$ site $l$ can be computed from the third
order spin dependent hopping perturbation. The general expression
is (where in the limit of small interactions $\epsilon$ is equal
to the difference in the $O$ and $Cu$ on-site energies)
\begin{equation}\label{eq:DM gener}
    \vec{D_{ij}}=\frac{8}{\epsilon^{3}}\sum_{l}(t_{il}t_{lj}+\vec{t}_{il}\vec{t}_{lj})\sum_{l}(t_{il}\vec{t_{lj}}-t_{jl}\vec{t_{il}}+\vec{t}_{il}\times\vec{t}_{lj})
\end{equation}

The non-zero value of $\vec{D_{ij}}$ comes from spatial inhomogeneity of (by its definition periodic) OCP pattern, which implies
$\vec{t_{lj}}\neq\vec{t_{il}}$. Taking into account $\vec{t}\sim
10^{-6}eV$ and using the fact that $\epsilon\leq t$ in $Cu$-based
ladders (and when $U\rightarrow 0$ as for the {\emph{C2S2} phase, $\epsilon$
determines inter-level distances), we can estimate $D\sim
10^{-5}eV$ (it is of order $\sim
\vec{t}(\frac{t}{\epsilon})^{3}$). This value is one order of
magnitude larger than magnetic dipole-dipole interactions expected
for such compounds (see Appendix).

We see that in the OCP case, where $\vec{t}\| z$, the
cross-term term in the second parenthesis of Eq.~(\ref{eq:DM
gener}) is zero, and hence $\vec{D}$ is perpendicular to the
plane of the ladder. It also implies that this four-fermion
operator will be proportional to the asymmetry between
neighboring $Cu$-$Cu$ links (\emph{il-th} and \emph{lj-th}). We
note that if the local magnetic moments were constant along the
ladder, $\vec{D}=0$.

Thus, the periodicity of the symmetry breaking perturbation plays
a critical role here. The OCP is superimposed on a DW
with a $(2k_{Fo})^{-1}$ periodicity in real space, and the
Fourier transform of $D({\mathbf x})$ -- which originates from the local
magnetic fields -- has components at $|q|=2k_{Fo}$ only, i.e
corresponds to a backward scattering vertex. This mirrors the fact
that the spin-rotational symmetry breaking physics brought about by
currents is due to the \emph{o}-band and can only involve carriers scattering
in this band.

When $\vec{D_{ij}}\|z$ one may deduce that the processes which
explicitly break spin-rotational symmetry are of the form
$\hat{S_{x}}\hat{S_{y}}$. Using spin-flip
operators we have terms like $S_{+}^{\dag}S_{-}+h.c.$. Physically
this indicates that spin is not a conserved quantity anymore and
that new instabilities may arise as a result of the additional
scattering channels. The new term in the hamiltonian is
\begin{equation}\label{eq:DMfermion}
    H_{D-M}=g_{fo}\int \psi_{Lo\sigma}^\dag \psi_{Ro\sigma}^\dag \psi_{Lo\bar{\sigma}}\psi_{Ro\bar{\sigma}}
\end{equation}
where $g_{fo}$ is the $2k_{Fo}$ component of the Fourier
transform of $D_{ij}$. This can be bosonized in the standard way:
\begin{equation}\label{eq:DMbozon}
    H_{D-M}=g_{fo}\int dr \cos(2\theta_{so})=g_{fo}\int dr\cos(\theta_{+s}+\theta_{-s})
\end{equation}
Since spin-flip terms are involved in the fermionic operators,
none of these will give simple density-density contributions to
the bosonized expression of the hamiltonian; instead they produce
non-linear interaction terms which require RG perturbative
treatment. They may thus open a gap in the spectrum and cause new
types of orderings in the ladder (which were previously forbidden
as they explicitly break SU(2) symmetry). We
investigate this possibility in the next section.

\section{Analysis of the SU(2) symmetry breaking terms using bosonization and RG arguments}\label{sec:results}

The standard way to deal with four fermion operators, which
produce cosine terms in bosonization language, is to follow an RG
approach. This was extensively discussed in the context of SU(2)
symmetric two-leg ladders (see for example Sec. III in
~\onlinecite{chudzinski_ladder_long}) and we transpose it to the present
situation.

As discussed in the previous section, we add the $g_{fo}\sim
D(2k_{Fo})$-type contributions to the Hubbard-type interactions
and,  following the standard procedure for cosine terms in the
bosonization representation, we perform an RG analysis of the
set of four fermions operators. Because spin-flip processes are
now allowed, we have to add scattering events where -- for
instance -- two incoming fermions are in a spin-up state and
the two outgoing particles are in the spin-down state.

\subsection{The extended system of RG equations}

We proceed along the same lines as in our previous studies of
two-leg $Cu$-$O$ ladders; however additional subtle factors
have to be taken into consideration. The fact that $D(q)$ has always just
one component in momentum space simplifies the analysis and
enables us to exclude some processes

One should take the system of RG equations derived for two-leg
$Cu$-$O$ Hubbard ladders ( Eq.~[17] to Eq.~[31] in Ref.~\onlinecite{chudzinski_ladder_long}) and  add to it two equations
describing the renormalization of the $g_{fo}$ terms (intra \emph{o}-band
spin-flip scattering):
\begin{multline}\label{eq:gf1}
    \frac{dg_{f1}}{dl}=g_{f1}\cdot
    (2-(K_{2}^{-1}+K_{1}^{-1}))+\\
P_{1}Q_{1}(K_{2}^{-1}-K_{1}^{-1})g_{f2}
\end{multline}
\begin{multline}\label{eq:gf2}
    \frac{dg_{f2}}{dl}=-g_{f2}\cdot
    (2-(K_{2}^{-1}+K_{1}^{-1}))+\\
P_{1}Q_{1}(K_{2}^{-1}-K_{1}^{-1})g_{f1}
\end{multline}
where
$g_{f1}=\frac{g_{fo}+g_{f\pi}}{2},g_{f2}=\frac{g_{fo}-g_{f\pi}}{2}$,
and by definition, in our case,  $g_{f\pi}\equiv 0$ initially.
We are using the same notation as in Ref.
~\onlinecite{chudzinski_ladder_long}: $K_{1,2}$ are the LL parameters of
the eigenmodes in the spin sector, i.e linear combinations of
$K_{s\pm}$ such that $K_{1}=P_{1}^{2}K_{s-}+Q_{1}^{2}K_{s+}$ and
$K_{2}=P_{1}^{2}K_{s+}-Q_{1}^{2}K_{s-}$. The initial values of
$K_{s\pm},~P$, and $Q$ are determined from the Fermi velocities in
the two bands and interactions of the forward type which are
included in the LL matrix $\hat{K}$ (note that
$P_{1}^{2}+Q_{1}^{2}=1$). Since $g_{fo}$ has a cosine form in
bosonization language, it does not change the initial values of
the entries of the LL matrix $\hat{K}$.

Basically the new couplings $g_{f1}$ and $g_{f2}$ can be treated
in a way similar to $g_{1}$, $g_{2}$; as they pertain to
the $\theta_{+s}$ field, these cannot couple with other
interactions $g_{i}$. The second order $g_{fo}^{2}$ terms will
enter the renormalization equations of the LL parameters $K_{i}$
and $\cot 2\alpha$ (eigenbasis angle) in the spin sector:

\begin{equation}\label{eq:Ks1}
    \Delta\frac{dK_{1}}{dl} =\frac{1}{2}(g_{f1}^{2}+g_{f2}^{2})+f(P_{1})(g_{f1}g_{f2})
\end{equation}

\begin{equation}\label{eq:Ks2}
    \Delta\frac{dK_{2}}{dl} = \frac{1}{2}(g_{f1}^{2}+g_{f2}^{2})-f(P_{1})(g_{f1}g_{f2})
\end{equation}

\begin{equation}\label{eq:alpha}
    \Delta \frac{dB_{12}}{dl} = h(P_{1})g_{f1}g_{f2}
\end{equation}

where, as before:
\begin{equation}
 f(P_{1})=(P_{1}Q_{1}+\frac{1}{4}\frac{P_{1}^{2}-Q_{1}^{2}}{P_{1}Q_{1}})^{-1}
\end{equation}
\begin{equation}
 h(P_{1})=((P_{1}Q_{1})^{2}+0.25(P_{1}^{2}-Q_{1}^{2}))^{-1}
\end{equation}

We notice that, as one might have expected, only the spin sector
is affected by the new non-linear terms. The modifications that
these terms cause to the RG flow (in particular as far as the
eigenbasis rotation is concerned) are not easy to evaluate, because -- as we argue below -- several
effects play a role. A qualitative analysis of the RG equations
only allows one to predict that the evolution of $\cot 2\alpha$ during
the flow will be much slower than in the standard case.

The initial values of the $K_{i}$ parameters will be different
from what they are in the standard case, if only because of
spin-charge mixing (following our discussion in Sec.
\ref{subsec:spindependenthopping}, spin up and down are not
degenerate any longer). This effect was partially accounted for by
Moroz et al.\cite{Moroz_SO}: they found that it can be implemented
through a proper change of the scaling dimension of $\cos(\phi)$
(Eq.~[38] in ~\onlinecite{Moroz_SO}). More explicitly, $K_{s+}$ and
$K_{c+}$ will mix (just as $K_{s\pm}$ do, in response to
particle-hole symmetry breaking) with a mixing parameter
proportional to
$\tilde{\epsilon}=(V_{F\uparrow}-V_{F\downarrow})/V_{F}$ 

Assuming that the \emph{o}-band is decoupled from the $\pi$-band, we can easily carry over Moroz's
results to the ladder case and conclude that the mixing of spin and charge
generated by the spin-orbit coupling is irrelevant in the
\emph{o}-OCP case, because here $\tilde{\epsilon}\ll
\tilde{\epsilon}_{crit}=\frac{V_{s+}^{0}}{V_{Fo}}$

Yet, another effect does play a role. As we pointed out in the
band structure discussion, because of the Fermi vector
mismatch, perpendicular spin scattering is reduced compared to
parallel spin scattering (as is the case with umklapp
scattering, slightly away from $k_{F}=\pi/2$). This causes a
shift of the phase boundaries in the ground state phase
diagram, but it does not generate any new phases. To
demonstrate this explicitly requires a full blown RG treatment
but  heuristic arguments pertaining to the RG flow in the
\emph{C2S2} phase -- presented in the next section -- give clear
indications of this trend.

\subsection{Relevance of the $g_{f}$ terms}

The overall structure of the phase diagram that emerges from the
RG analysis, when these additional terms are included, is similar
to that obtained in our previous papers, so we discuss here the
stability of the phases that were found for ladders possessing
SU(2) symmetry. Let us start from the half filled case when there
is on average one carrier per copper atom and move towards the
highly doped regime when the $\pi$-band is nearly empty. 

The low doping \textbf{C1S0 phase} is not affected, since the
\emph{o}-OCP does not exist, and hence $D(2k_{Fo})=0$. However, to
be on the safe side, we have checked that even if $D(2k_{Fo})\neq
0$ interactions of the $g_{fo}$ type are irrelevant. This can be
justified by symmetry considerations regarding the eigenbasis: in
the low doping regime $B_{+/-}$ is the proper basis of LL modes,
thus a small $g_{fo}$, which induce instability in the $B_{o\pi}$
basis cannot be relevant in this doping range. It must be
interband spin flip scattering which causes a relevant
perturbation of the \textbf{C1S0 phase} -- the possibility of the
other types of spin flip interactions are briefly discussed in the
Appendix.

A significant difference, caused by $g_{fo}$, is found for the
massless \textbf{C2S2 phase} that was present in a finite,
intermediate doping range of the SU(2) symmetric problem. We
recall, that for this phase, $B_{o\pi}$ is the proper basis of the
spin sector, thus interband instabilities are suppressed. Adding
$g_{fo}$ favors a flow towards this basis. We can simplify our
analysis even further if we assume that in given range of dopings,
the RG equations describing the evolution of the \emph{o}-band are
decoupled from the rest of RG system.  This allows us to work with
Eq.(\ref{eq:gf1}) and the equation (derived previously
\cite{chudzinski_ladder_long}) describing intraband
backscattering:
$$
\frac{dg}{dl}=g\cdot (2-(K_{2}+K_{1}))+ P_{1}Q_{1}(K_{2}-K_{1})g
$$
$g_{o}$ was proved to be irrelevant since the initial $K_{2}>1$
and we would need a large value of $Q_{1}^{2}$ in order to open a
gap in $\phi_{so}$. Roughly speaking $K_{so}$ was approaching its
fixed point $K_{so}\rightarrow 1$ too slowly and the flow was
finally only marginal. In our extended case $g_{fo}$ tends to
increase $K_{so}$ or at least (in the beginning) protects its
constant value. The arguments given in
~\onlinecite{chudzinski_ladder_long} work exactly in the opposite way for
the $\cos\theta_{so}$ instability, which is relevant provided
$K_{so}>1$. This can be quantified by studying an invariant of our
extended RG system:
\begin{equation}\label{eq:RGinvariant}
    A^{2}=(K_{so}-1)^{2}-(g_{o}^{2}+g_{fo}^{2})
\end{equation}
We see that by adding the new terms $A^2$ is shifted (from
$A\approx 0$) to negative values; it is known that an imaginary
$A$ in the Kosterlitz-Thouless RG flow corresponds to a
divergent trajectory that is to a relevant instability destroying the LL fixed
point. The only possibility corresponding to that case is that $g_{fo}$
becomes the relevant perturbation.

Our earlier observation that momentum mismatch reduces the initial
value of the perpendicular spin scattering interaction while
leaving those of the parallel and spin flip processes unchanged
reinforces the inequality $(K_{so}-1)>g_{\perp}$. We then study a
Kosterlitz-Thouless RG flow with all the $g_{\perp}$ terms smaller
than  the $g_{\parallel o}$ terms; in that case we find that we
move away from the SU(2) separatrix and scale towards larger
$K_{so}$, which favors the new intermediate phase (with the
$\theta$ field ordered).

One finds that a new gap opens in the spectrum of the
$\theta_{so}$ field. We find that $C2S2\rightarrow C2S1'$, with an
ordering within the \emph{o}-band. Using a bosonic version of
order operators we find that SDW in the plane of the ladder
($SDW_{x}$ in the \emph{o}-band) is favored. The order operator is
defined as follows:
\begin{align*}\label{eq:SDWorder}
    O^{x}_{SDWo}(x) &=
    \psi^{\dag}_{oR\uparrow}\psi_{oL\downarrow}(x)+\psi^{\dag}_{oR\downarrow}\psi_{oL\uparrow}(x)\\
    &\simeq \exp(-2\imath k_{Fo}x)\exp(\imath\sqrt{2}\phi_{co}(x))\cos(\sqrt{2}\theta_{so}(x))
\end{align*}
The interaction breaking spin-rotational SU(2) symmetry dominates
in this regime, so it is clear that the resulting phase also
breaks this symmetry.

This simplified analysis is confirmed when we perform the actual
RG calculation including the additional terms. Furthermore, we are
able to obtain the exact boundary between the $g_{fo}$ and $g_{o}$
dominated phases (i.e the boundary between the SU(2) symmetric and
symmetry broken phases). We also see that the assumption
$K_{so}\simeq const$ is justified, so we can approximate the value
of a gap by taking the following RG flow: $\frac{dK_{so}}{dl}=0$,
from which one deduces that a proper estimate of the gap value is
\cite{giamarchi_book_1d}:
\begin{equation}\label{eq:gap}
    \Delta\simeq\Lambda_{0}g_{fo}^{\frac{1}{2-2K_{so}^{-1}}}
\end{equation}
Assuming $K_{so}\leq 2$ (strong, but local, Hubbard interaction
limit) we find $\Delta\simeq 0.01 D$, which gives quite a small
but experimentally accessible transition temperature $\in
(1,10)mK$.

For the \textbf{C2S1 phase} the additional $g_{fo}$ terms compete
with the interaction $g_{o}=g_{1}+g_{2}$, so the theory of spin
flip effects in the case of a single chain applies. The reasoning
is very similar to that for the \textbf{C2S2 phase}, except that
now $g_{o}$ dominates. To put it on a more quantitative level: as
long as the initial $g_{f}$ is smaller than $g_{o}$ -- which is a
reasonable assumption, because $g_{o}$ is a relevant perturbation
dominant in this doping range -- the ordering of the $\phi_{so}$
field overwhelms spin-flip processes. Giamarchi and Schulz have
found that the critical line for this problem is at $g_{f}=g_{o}$
\cite{giamarchi_spin_flop}. This assumption can fail only close to
the low doping boundary of the \emph{C2S1} phase, since the
strength of $g_{o}$ decreases as one reaches $\delta_{c2}$. We can
thus conclude that, except for a shift of the lower boundary, the
CDW state is unaffected by the spin-orbit coupling. The new phase
diagram, in the presence of $g_{fo}$ is summarized in the figure
below (Fig.\ref{fig:phase_diag_modif}).

\begin{figure}
  \centering
  \includegraphics[width=0.45\textwidth]{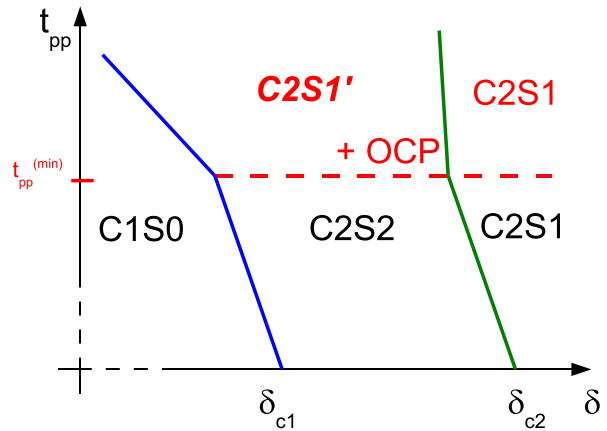}\\
  \caption{Modified phase diagram (to be compared with Fig.\ref{fig:dopin}) of the two-leg Cu-O ladder when intra \emph{o}-band spin-flip processes are included: as before $\delta$ is the hole doping and
  $t_{pp}$ inter-oxygen hopping. In the intermediate doping range -- where a \emph{C2S2} phase was previously found in the SU(2) symmetric problem, a \emph{C2S1'} state is present, which has a SDW in the plane of the ladder as a result of the opening of a spin gap.}\label{fig:phase_diag_modif}
\end{figure}

\section{Discussion}\label{sec:discussion}

The breaking of SU(2) symmetry, that occurs in the presence of the
OCP, opens a gap in the spin mode of the \emph{o}-band. In the
range of dopings where previously a critical line was found, the
\textbf{C2S2} phase is  replaced by a different ground state,
namely a phase with in-plane $SDW_{x}$ denoted by \textbf{C2S1'}.
An important question is whether this unusual phase is detectable,
or if any small symmetry breaking will make it undistinguishable
from the standard, high doping \textbf{C2S1} phase. To answer this
question we present a discussion along three lines: the first one
is devoted to high temperature correlation functions, the second
to transport properties and the third to magnetic
properties. A limitation of the
discussion presented below, which affects experiments, is the energy scale: the
effects will be clearly visible only in a temperatures range below
the gap $\Delta$.

One may also wonder whether the existence of the OCP is the
only source of SU(2) symmetry breaking. As the $Cu$-$O$ lattice
forms a fairly rigid structure, it is stable both against
crystal field $d$ levels splitting and standard $L-S$ coupling
$\sim Z^{2}$  on heavy $Cu$ ions (with $Z=29$), and, insofar as
we may neglect size effects, the presence of an OCP is the
source of symmetry breaking. A detailed discussion of these
issues will be presented in the Appendix.

\subsection{Correlation functions for the LL phase}

Correlation functions for LL with spin-orbit interactions have
been evaluated both for two- \cite{Moroz_SO} and four- fermion
\cite{Iucci_SO} operators. Let us briefly summarize here their
main features. Chiral separation is still preserved and
correlators can still be factorized in terms of two modes, but not
spin and charge anymore. Interactions and spin orbit coupling
(L-S) reinforce one another; in the presence of L-S, the density
of states decays faster than for free fermions, and the difference
between the velocity of the two modes $\delta V=
V_{\uparrow}-V_{\downarrow}$ is proportional to $\vec{t}$.
According to \cite{Iucci_SO} the exponents are slightly changed
when a term proportional to a difference of the velocities is
added (we use these results slightly modified, because it was
Rashba-type spin-orbit coupling treated in Ref.~\onlinecite{Iucci_SO}, while
we are interested in Zeeman-type coupling); for example the
behavior of our dominant fluctuation is given by (at zero
temperature, up to first order in $\delta V$):
\begin{equation}\label{eq:SDW_exp}
O^{x}_{SDWo}(r)\simeq\cos(2k_{Fo}x)r^{-(K_{co}+1/K_{so}-\theta^{s})}
\end{equation}
where $\theta^{s}\sim \delta V$. As a consequence the four fermion
operators have a non-zero conformal spin proportional to $\delta
V$.

In our case intra- \emph{o}-band L-S coupling is quite weak (this
small value enabled us to neglect some effects of spin-charge
mixing during the RG flow). Accordingly, in the high temperature
LL phase, the functional (power law) form of the correlators will
be the same for the \emph{C2S1} and \emph{C2S1'} phases, and the
only difference comes from small changes in the exponents. Taking
into account small (and oscillating) value of $\delta k$ ($\delta
V$), and experimental difficulty with extracting exact value of
bare LL exponent $K_{io}$, we doubt that \emph{C2S1'} existence
can be shown in this way.

More visible signatures can be expected at low temperature where
the $\theta_{so}$ field is ordered, as all the observables
depending on magnetic susceptibilities -- such as NMR responses --
are affected (notwithstanding the relatively smallness of the
gap): if the $\theta_{so}$ field is locked, then averages over the
$\phi_{so}$ field go abruptly to zero. According to
\cite{chudzinski_ladder_long} one expects quite abrupt suppression
of the \emph{o-band} electronic part of the Knight shift $K$ and
relaxation rate $(1/T_{1})$ at $T<\Delta$ (for example $K\sim
\langle\cos\phi_{so}\rangle$).

\subsection{Transport properties: analysis of the impurity problem}

Let us briefly discuss here what happens if a non-magnetic
impurity is introduced in the two-leg $Cu$-$O$ ladder. This
problem was discussed in detail recently \cite{chudzinski_impur};
here we focus on the consequences of spin-rotational symmetry
breaking. Specifically, we investigate
 the doping range where the
\emph{C2S2} phase was stable in the absence of spin-orbit
scattering. In the previous section we have shown that $K_{so}$
behaves differently at the lowest temperatures, because of a gap
opening due to spin-orbit terms. So the temperature dependence of
 the transmission through a single impurity -- which is connected to the
renormalized backscattering $V_{o}[l;K]$ -- will also be affected.

Spin-orbit coupling is present only inside the \emph{o}-band, so we may
restrict our analysis to reflection (transmission) of carriers
inside this band. In such a simplified case, the RG
equation for backscattering on the impurity is given by
\begin{equation}\label{eq:RGimp}
    \frac{dV_{o}}{dl}=(2-K_{so}-K_{co})V_{o}
\end{equation}
instead of Eq.~[15a] in Ref.~\onlinecite{chudzinski_impur}, where it was assumed that
$K_{so}=1$). From this, it is clear that increasing $K_{so}$
(which promotes an ordering of the $\theta_{so}$ field) can
cause $V_{o}$ to become irrelevant. More precisely, this
operator is always relevant at the beginning of its flow,
because $K_{co}<1$ (and this value is constant in the absence
of umklapp terms), but this trend may change as the energy is
decreased. This would give rise to an unusual temperature
dependence of the conductivity.

On the other hand, if the initial value of $K_{co}$ is
sufficiently small (strongly interacting case), we would arrive at
an open boundary problem ($V_{o}$ relevant). In our previous paper
\cite{chudzinski_impur}, intraband Kondo physics was suggested for
this case. The following RG equation describes this regime
(assuming open boundary conditions (OBC))

\begin{equation}\label{eq:Jk}
    \frac{dJ_{o}}{dl}=J_{o}(1-K_{so}^{-1})
\end{equation}
We see here that if SU(2) symmetry is broken by the \emph{o}-OCP,
this coupling becomes relevant, which confirms our previous
predictions. In the limit of quite strong Hubbard interactions
$K_{co}\rightarrow 1/2$ -- which is the case when $K_{so}\gg 1$ --
it can give rise to the unusual regime predicted many years ago by
Furusaki and Nagaosa \cite{Furusaki_RG2} who took second order
tunneling processes into account, showing that double spinon
tunneling events may dominate the physics around the impurity. The
competition between OBC Kondo physics and spinon transmission will
take place in the strong interaction limit (the barrier asymmetry
will also play a significant role), but this is beyond the scope
of the present paper.

An experimental consequence of this effect, is that in the phase
where spin-orbit coupling dominates, the usual line broadening
caused by impurities that is expected in an NMR measurement, will
not be seen. This is because spinons can be transmitted through
non-magnetic impurities (and hence the backscattering in the spin
sector is expected to be weak).

\subsection{Magnetic properties}

Obviously, the local magnetic field will affect the Knight shifts
of the NMR lines. The magnetic field is alternating and
incommensurate with the lattice: as a result of such local
magnetic field variation NMR lines will not have standard (thermal
broadened) shape, but it will depend on the number of atoms
feeling a given value of the field $N(B_{loc})$. The
incommensurability with the lattice allows us to rephrase this
question and ask what is the density of states for a cosine curve.
Thus the NMR line-shape resemble 1D density of states with two
peaks line-shape (1D density of states has $1/\omega$ singularity
coming from the bottom of the cosine band). The splitting is equal
to $2B_{max}$ and, because the effective field is different at
each atom inside unit cell, thus different line-shapes are
expected: the widest on on-leg oxygen atom, and non-perturbed
(zero split) on the on-rung, central oxygen. This effect should be
distinguished from the one described in the previous paragraph, in
total OCP modifies NMR lines in two ways: it suppress the
amplitude of the satellites, of the impurity at the origin, and
splits the central peak. Unfortunately estimated values of this
splitting are at the limit of the current experimental resolution,
so this primary effect might be difficult to detect. One can
alternatively detect anisotropic g-factors through Zeeman effect.

There will also be a signal coming from the magnetic ordering
itself. We predict that there is a magnetic moment component
perpendicular to the ladder plane due to $B_{eff}$ induced by the
\emph{o}-OCP (primary effect), but there is also a secondary
effect: from our RG procedure we have found an in-plane component
corresponding to a $SDW_{x}$, with the same periodicity as for the
staggered moments along $z$. The fact that both components have
the same periodicity in $k$ space, $2k_{Fo}$, is quite important:
they can add up creating a moment tilted out of plane. The tilting
angle will be determined by their relative amplitudes. We have
started with $\mu_{z}=0.1\mu_{B}$ which induce an ordering gap
$\Delta$. The value of $\mu_{x}$ is connected with the average
$\mu_{x}\sim\langle\cos\theta_{so}\rangle$ so it will be reduced
with increasing temperature. Even at zero temperature quantum
fluctuations cause $\langle\cos\theta_{so}\rangle\neq 1$. In order
to estimate the average at zero temperature we compare the values
of the gap we found in Sec.\ref{sec:results} with the one found
when the LL velocity (kinetic energy) goes to zero
$\Delta_{\infty}$. It is known that $\Delta_{\infty}\simeq
|\vec{D}|$. The ratio $\Delta/\Delta_{\infty}=0.01$ is
proportional to $\langle\cos\theta_{so}\rangle$ and allows us to
estimate $\mu_{x}\approx 0.01 \mu_{B}$, from which we deduce a
tilting angle of the order of $\vartheta\sim 6deg$ (counted from
vertical axis). This kind of modulated (in space) magnetic moments
composition should be detectable using elastic neutron scattering,
but quite low temperature ($\sim 10mK$) would be required for this
experiment.

\section{Conclusions}

The main question which we addressed in this work was: what is
the feedback effect of an OCP  on
the properties of the ladder, if one takes into account
the coupling between the electrons and the induced orbital
moment? Such feedback can in fact potentially provide an
additional method to probe the existence of the OCP; it
does not require a direct measurement of the induced -- and
quite tiny-- magnetic field. Detecting such feedback can however prove challenging,
due to two important constraints. First, one would need to
obtain highly doped ($\sim 25\%$) two-leg ladders; second,
experiment have to be performed at temperatures below $\Delta$.
Still, we hope that our present detailed insight into the low
energy properties of the intermediate phase will invite
experimental and numerical investigations.

A non-trivial issue is how reliable our value of $\Delta$ is. We
have used a weak coupling RG approach which is reasonable in the
case of the otherwise gapless $C2S2$ phase: the statement
$\Delta<|\vec{D}|$ is justified. Thus, in our framework the only
way to increase $\Delta$, assuming a given amplitude of the OCP
($\sim\mu_{z}$) is by taking larger $\lambda$ coefficient in
Eq.(\ref{eq:hopp}) (this would give larger $|\vec{t}|$). This may
arise because of particular, unknown property of copper oxides
and/or constructive interference of L-S couplings defined on the
top of the DW. Larger gap value will push the tilt angle
$\vartheta$ towards larger values.

Beyond the experimental manifestations, which we have discussed in
Sec.\ref{sec:discussion}, the magnetic moment pattern should be
detectable using elastic neutron scattering. Such
experiments were recently performed in the pseudogap phase of 2D
HTSC \cite{BFauque_neutrons, Baledent_OCP}. The data was interpreted as evidence for out-of plane magnetic
moments with an alternating tilt. There is a similarity with our
prediction, but one must not forget that there are several
important differences between the OCP that we have discussed and
the one that could serve to interpret the experiments in cuprates:
in our 1D case the pattern is modulated in space with a $k_{Fo}$
periodicity (this modulation was in fact critical to obtain a
non-zero coupling with the spin degree of freedom) and has a
different internal symmetry: $\theta_{1}$ instead of $\theta_{2}$
in Varma's notation \cite{CVarma_orbitalcurrents}. One must keep
in mind these caveats, nonetheless our result invites one to look
for similar effects in two-leg ladders. 

To summarize: despite certain similarities, the 2D cuprates studied
in Ref.~\onlinecite{BFauque_neutrons, Baledent_OCP} cannot be treated as
an experimental realization of our problem. A good candidate would
be a doped ladder, such as the so called telephone compound ($Sr_{14-x}Ca_{x}Cu_{24}O_{41}$),
provided sufficient doping levels can be attained (from the standpoint of solid-state chemistry,
this is a highly non-trivial task). Neutron diffraction or $\mu$SR experiments would be desirable for such a ladder system.

An important point is the possible role of apical oxygens.
In the framework of a tight binding model the new hoppings will
make the two bands ($o$ and $\pi$) more asymmetric; this implies
larger effective $t_{pp}$, which shifts the \emph{C2S2} phase towards
smaller dopings $\delta$ (see Fig.\ref{fig:dopin}). New conducting
paths will also increase the OCP amplitude for a given value of $t_{pp}$,
shifting  the critical $t_{pp}^{min}$ to smaller values. As a result the new
phase \emph{C2S1'} will be easier to access experimentally in the
$(\delta-\frac{t_{pp}}{t_{pd}})$ plane. The new current paths can
also give rise to higher order L-S couplings: for example
$\vec{\mu}$ can induce additional currents flowing in the loops
involving the apical oxygen. We may conclude that including apical
oxygen should favor the effects predicted in this paper.

This statement, about a positive role of apical oxygens, is in qualitative
agreement with recent work \cite{Bulaevskii_added} studying currents emerging because
of particular spin-textures.

A non-trivial outcome of this work is that it underscores the
hidden connection between a phase with OCP and a pseudo (partial)
gap in the spin sector. The emergence of such a gap  shields the
intermediate, previously marginal, phase from weak perturbations
such as interladder hopping or disorder. In conclusion our work
shows that a strong perturbation is caused by the presence of the OCP
and validates the phase diagram for two leg Hubbard ladders which had
been obtained in Ref.~\onlinecite{chudzinski_ladder_rapid}.

\section{Appendix: Other SU(2) symmetry breaking mechanisms}

In this Appendix, for the sake of completeness, we briefly discuss
alternative sources of spin-rotational symmetry breaking. A few of
them come readily to mind: dipole-dipole interactions, standard
$\vec{L}\cdot\vec{S}$ spin-orbit coupling and Rashba mechanism.

\textbf{1)} Direct spin-dipolar interactions. The value of this
spin anisotropy depends on the average distance between carriers
as $\sim 1/\langle r\rangle^{3}$. The $\langle r\rangle$ is
obviously changing when one add carriers into the ladder which
implies a doping dependence. We can use values found for the half
filled case in a similar compound \cite{Hanzawa_dipolar_int} and
simply assume that carriers are more diluted. For the most
interesting phase (\emph{C2S2}) there are at least two times less
carriers which gives reduction by a factor 8, thus we have
estimated the dipole-dipole anisotropy to be one order of
magnitude smaller than OCP effects.

Besides, in the \emph{C2S2} regime, OCP effects are predominant
because they possess the proper \emph{o-band} symmetry, in
contrast with the generic form of dipolar interactions. Thus we
assume that this mechanism cannot compete nor mimic the OCP
effects predicted in this paper.

\textbf{2)} The spin orbit  $\vec{L}\cdot\vec{S}$ coupling is a
fairly typical effect for heavy atoms. Thus we should check
whether it could induce much stronger effects than those predicted
in our analysis.

The key point is the symmetry of the lattice: so long as copper
atoms are sitting in a $C_{4}$ symmetric environment, hoppings to
and from all the neighboring oxygens will conspire  to give a zero
net outcome. We can also invert this argument and say that if
electron ordering finds a way to lower the rotational symmetry of
a lattice then it would provide an efficient way to increase the
$\lambda$ coefficient in Eq.\ref{eq:hopp} and thus enhance effects
discussed in this paper. This property does not hold if the
lattice is distorted (in the Peierls or Jahn-Teller cases for
instance) or warped. As the lattice is quite rigid (and we are not
working at half filling) for ladders, these effects are not
expected to be significant in the bulk. The situation may be
different at an interface, but we are not dealing with such a case
in this paper. This is the reason why we focused on OCP-induced
effects.

The only exception  might be if the primary OCP involved apical
oxygens. Let us make a \emph{gedanken} experiment and assume that
there is a current flowing through some of bonds toward these
atoms. This might enhance the total amplitude of the OCP, lowering
the value $t_{pp}^{(crit)}$. It can also be a higher order
perturbation induced by the in-plane $SDW(x)$. The conducting
bonds tend to be shorter which can displace apical oxygens. The
emergence of such distortion will lower the lattice symmetry from
$C_{4}$ to $C_{2}$ allowing for the non-zero net expectation value
of
%$\vec{L}\cdot\vec{S}$
 the spin-orbit coupling on $Cu$ atoms. This would
cause a $S^z$ dependent hopping, as can be seen from
Eq.~[\ref{eq:hopp}] which shows that the displacement is
equivalent to admixing $d_{x^{2}-y^{2}}$ orbitals with the apical
oxygen band. Importantly, this effect is caused by the OCP, thus
it should have the same periodicity and phase. This additional L-S
coupling enhances the OCP induced effects discussed above.

\textbf{2)} Another possible mechanism of interest
in the cuprate case, giving spin-dependent hopping, arises from the
Rashba effect which couples directly electron motions to their
spins. It describes the precession of carriers in the presence of a
strong electric field. The Rashba Hamiltonian is:
\begin{equation}\label{eq:Rashba}
    H_{R}= \alpha (\vec{k}\times\vec{n})\cdot \vec{\sigma}
\end{equation}
where $\alpha$ is a coupling constant and $\vec{n}$ is a unit
vector in the electric field direction.
%(perpendicular to the plane in our case).
 Obviously, this coupling (which depends on the total
momentum of the carrier) will be different for the o and $\pi$
bands, because of the inequality $k_{Fo}>k_{F\pi}$ (one should not
forget that $k_{F\pi}$ has large on-rung components).

Firstly, let us consider the bulk of a ladder material. The
electric field which we expect will lie in the plane of the
ladders in a direction perpendicular to the legs. It will result
from the arrangements of DW in neighboring ladders (so again it
will be restricted to the high doping regime, and again have a
$(2k_{Fo})^{-1}$ periodicity in real space). There are two
limitations here: these arrangements can only be stable for
certain dopings (quarter-filling) and they are expected to be
quite small in amplitude.

Straightforward analysis (with again crucial  element of DW
modulation) of Eq.~[\ref{eq:Rashba}] gives a spin-dependent
hopping parameter $\vec{t}\sim [0, 0, t']$, so again it has only a
z-component. It implies that the analysis done in the core of this
paper applies also in this case. Thus, this kind of Rashba (if
such kind of perpendicular arrangement does exist) would enhance
the effects predicted before. The only difference which would
allow us to distinguish this case from that induced by an OCP is
the magnetization component perpendicular to the ladder plane.

Let us finish with a brief discussion of an interface where the
Rashba mechanism can induce quite new physics. In this case the
electric field will be perpendicular to the ladder plane, because
it arises from charge polarization effects at the interface. On
the interface, when charge differences are very high (polarization
catastrophe) at very small (interatomic) distances, this effect
can be really huge. What is more, in contradistinction with the
mechanisms that we discussed until now, the amplitude is not
alternating in space (the sign of the coupling is not
oscillating).

In the case of strong electric field there are only non-diagonal
(spin-flip) hoppings; level splitting is induced by the Rashba
term Eq.~[\ref{eq:Rashba}] (if the electric field $\vec{E}$ is in
the $z$ direction, $H_{R}$ can be expressed as $\sim [t', 0,
0]\cdot \vec{\sigma}$ in the \emph{o}-band and approximately by $
[0, t', 0] \cdot \vec{\sigma}$ in the $\pi$-band). In this case
spin and charge are strongly mixed thus spin-up and spin-down are
not a good quantum number any longer.

Also spin-flip interactions will be highly non-trivial in such a
model: we may find $\vec{t_{ij}}$ with perpendicular orientations
for instance when the two scattering carriers are from two
different bands. Thus it is the last term in the parenthesis of
Eq.~(\ref{eq:DM gener}) which is non-zero; we are then dealing
with interband scattering. We note that both $\vec{t}$ lie in the
ladder plane, so that, once again, $\vec{D_{ij}}\|z$ (spin-flip
process). We expect very different physics emerging in this case,
which is out of scope of this paper. A detailed RG analysis will
be postponed to  a future publication devoted to surface effects.

\bibliographystyle{prsty}
%\bibliography{totphys,ladder3b,impurities}

\begin{thebibliography}{10}

\bibitem{Alloul_YBCO}
H. Alloul, T. Ohno, and P. Mendels, Phys. Rev. Lett. {\bf 63},  1700  (1989).

\bibitem{PALeeetal_review}
P. Lee, N. Nagaosa, and X. Wen, Rev. Mod. Phys. {\bf 78},  17  (2006).

\bibitem{lederer_superconductivity_flux_ref}
P. Lederer, D. Poilblanc, and T.~M. Rice, Phys. Rev. Lett. {\bf 63},  1519
  (1989).

\bibitem{kotliar_liu_dwave_slavebosons}
G. Kotliar and J. Liu, Phys. Rev. B {\bf 38},  5142(R)  (1988).

\bibitem{chakravarty_ddw_pseudogap}
S. Chakravarty, R.~B. Laughlin, D.~K. Morr, and C. Nayak, Phys. Rev. B {\bf
  63},  094503  (2001).

\bibitem{affleck_marston}
I. Affleck and J.~B. Marston, Phys. Rev. B {\bf 37},  3774  (1988).

\bibitem{senthil_09}
T. Senthil and P.~A. Lee, Physical Review Letters {\bf 103},  076402  (2009).

\bibitem{gabay_RMP}
H. Alloul, J. Bobroff, M. Gabay, and P.~J. Hirschfeld, Reviews of Modern
  Physics {\bf 81},  45  (2009).

\bibitem{Kivelson_RMP}
S.~A. Kivelson, I.~P. Bindloss, E. Fradkin, V. Oganesyan, J.~M. Tranquada, A.
  Kapitulnik, and C. Howald, Rev. Mod. Phys. {\bf 75},  1201  (2003).

\bibitem{varma_3band_model}
C.~M. Varma, Phys. Rev. B {\bf 55},  14554  (1997).

\bibitem{CVarma_orbitalcurrents}
C. Varma, Phys. Rev. B {\bf 73},  155113  (2006).

\bibitem{BFauque_neutrons}
B. Fauqu{\'e}, Y. Sidis, V. Hinkov, S. Pailh{\`e}s, C. Lin, X. Chaud, and P.
  Bourges, Phys. Rev. Lett. {\bf 96},  197001  (2006).

\bibitem{Xia_optOAF}
J. Xia, E. Schemm, G. Deutscher, S. Kivelson, D. Bonn, W. Hardy, R. Liang, W.
  Siemons, G. Koster, M. Fejer, and A. Kapitulnik, Phys. Rev. Lett. {\bf 100},
  127002  (2008).

\bibitem{Kapitulnik_OCP}
A. Kapitulnik, J. Xia, E. Schemm, and A. Palevski, New Journal of Physics {\bf
  11},  055060 (18pp)  (2009).

\bibitem{Baledent_OCP}
Y. Li, V. Baledent, N. Barisic, Y. Cho, B. Fauque, Y. Sidis, G. Yu, X. Zhao,
  P. Bourges, and M. Greven, Nature {\bf 455},  372  (2008).

\bibitem{Sonier_OCPneutron}
J.~E. Sonier, V. Pacradouni, S.~A. Sabok-Sayr, W.~N. Hardy, D.~A. Bonn, R.
  Liang, and H.~A. Mook, Physical Review Letters {\bf 103},  167002  (2009).

\bibitem{strassle_lackOCP}
S. Str\"{a}ssle, J. Roos, M. Mali, H. Keller, and T. Ohno, Physical Review
  Letters {\bf 101},  237001  (2008).

\bibitem{weber_OCP}
C. Weber, A. L\"{a}uchli, F. Mila, and T. Giamarchi, Physical Review Letters
  {\bf 102},  017005  (2009).

\bibitem{greiter_exactdiag_currents}
M. Greiter and R. Thomale, Physical Review Letters {\bf 99},  027005  (2007).

\bibitem{thomale_exactdiag_currents}
R. Thomale and M. Greiter, Phys. Rev. B {\bf 77},  094511  (2008).

\bibitem{Bulaevskii_added}
L. N. Bulaevskii,  C. D. Batista, M. V. Mostovoy, and D. I. Khomskii, 
  Phys. Rev. B {\bf 78}, 024402 (2008).

\bibitem{piskunov_ladder_nmr}
Y. Piskunov, D. J{\'e}rome, P. Auban-Senzier, P. Wzietek, U. Ammerahl, G.
  Dhalenne, and A. Revcolevschi, Eur. Phys. J. B {\bf 13},  417  (2000).
  
\bibitem{piskunov04_sr14cu24o41_nmr}
Y. Piskunov, D. J{\'e}rome, P. Auban-Senzier, P. Wzietek, and A. Yakubovsky,
  Phys. Rev. B {\bf 69},  14510  (2004).

\bibitem{fujiwara03_ladder_supra}
N. Fujiwara, N. {M\=ori}, Y. Uwatoko, T. Matsumoto, N. Motoyama, , and S.
  Uchida, Phys. Rev. Lett. {\bf 90},  137001  (2003).

\bibitem{imai_NMR_doped_2ladder}
T. Imai, K. Thurber, K. Shen, A.W.Hunt, and F. Chou, Phys. Rev. Lett. {\bf 81},
   220  (1998).

\bibitem{kumagai_NMR_2ladder}
K. Kumagai, S. Tsuji, M. Kato, and Y. Koike, Phys. Rev. Lett. {\bf 78},  1992
  (1997).

\bibitem{orignac_2chain_long}
E. Orignac and T. Giamarchi, Phys. Rev. B {\bf 56},  7167  (1997).

\bibitem{schollwock_CDW+current}
U. Schollwock, S. Chakravarty, J. Fjarestad, J.~B. Marston, and M. Troyer,
  Phys. Rev. Lett. {\bf 90},  186401  (2003).

\bibitem{chudzinski_ladder_long}
P. Chudzinski, M. Gabay, and T. Giamarchi, Phys. Rev. B {\bf 78}, 075124 2008.

\bibitem{chudzinski_ladder_rapid}
P. Chudzinski, M. Gabay, and T. Giamarchi, Phys. Rev. B {\bf 76}, 161101 (R)
  (2007).

\bibitem{giamarchi_book_1d}
T. Giamarchi, {\em Quantum Physics in One Dimension} (Oxford University Press,
  Oxford, 2004).

\bibitem{voit_bosonization_revue}
J. Voit, Rep. Prog. Phys. {\bf 58},  977  (1995).

\bibitem{balents_2ch}
L. Balents and M.~P.~A. Fisher, Phys. Rev. B {\bf 53},  12133  (1996).

\bibitem{Moroz_SO}
A.~V. Moroz, K.~V. Samokhin, and C.~H.~W. Barnes, Phys. Rev. B {\bf 62},  16900
   (2000).

\bibitem{giamarchi_spin_flop}
T. Giamarchi and H.~J. Schulz, J. Phys. (Paris) {\bf 49},  819  (1988).

\bibitem{Iucci_SO}
A. Iucci, Phys. Rev. B {\bf 68},  075107  (2003).

\bibitem{chudzinski_impur}
P. Chudzinski, M. Gabay, and T. Giamarchi, NJP {\bf 11},  055059  (2009).

\bibitem{Furusaki_RG2}
A. Furusaki and N. Nagaosa, Phys. Rev. B {\bf 47},  4631  (1993).

\bibitem{Hanzawa_dipolar_int}
K. Hanzawa, Journal of the Physical Society of Japan {\bf 63},  264  (1994).

\end{thebibliography}

\end{document}